\documentclass{article}
\usepackage{amsmath,amsfonts,amssymb,bm,mathrsfs}
\usepackage{graphicx,color,rotating}
\definecolor{MidRed}{rgb}{0.78,0,0}
\definecolor{MidGreen}{rgb}{0,0.65,0}
\definecolor{MidBlue}{rgb}{0,0,0.68}
\newcommand{\unit}[1]{\ensuremath{\mathrm{#1}}}
\newcommand{\ohm}{\ensuremath{\mathrm{\Omega}}}

\newcommand{\req}[1]{(\ref{#1})}

\title{Phase Noise and Jitter in Digital Electronics}
\author{Claudio E. Calosso\thanks{CEC is with INRIM, 
Division of Physics Metrology, Torino, Italy.  E-mail c.calosso@inrim.it} and Enrico Rubiola\thanks{ER is with FEMTO-ST Institute, Univ.\ Bourgogne - Franche Comt\'{e}, and CNRS.  Address: ENSMM, 26 Chemin de l'Epitaphe, 25030 Besan\c{c}on, France.  Home page http://rubiola.org.  E-mail rubiola@femto-st.fr}}
\date{\today}
\def\myheaders{\sffamily\small C.\,E.\,Calosso and E.\,Rubiola\hfil Phase Noise and Jitter in Digital Electronics \hfil}
\begin{document}
\maketitle

\begin{abstract}
This article explains phase noise, jitter, and some slower phenomena in digital integrated circuits, focusing on high-demanding, noise-critical applications.
We introduce the concept of \emph{phase type} and \emph{time type} (for short, $\varphi$-type and $\mathsf{x}$-type) phase noise.  The rules for scaling the noise with frequency are chiefly determined by the spectral properties of these two basic types, by the aliasing phenomenon, and by the input and output circuits.

Then, we discuss the parameter extraction from experimental data and we report on the measured phase noise in some selected devices of different node size and complexity.  
We observed flicker noise between $-80$ and $-130$ \unit{dBrad^2/Hz} at 1 Hz offset, and white noise down to $-165$ \unit{dBrad^2/Hz} in some fortunate cases and using the appropriate tricks.

It turns out that flicker noise is proportional to the reciprocal of the volume of the transistor. This unpleasant conclusion is supported by a gedanken experiment.

Further experiments provide understanding on: (i) the interplay between noise sources in the internal PLL, often present in FPGAs;  (ii) the chattering phenomenon, which consists in multiple bouncing at transitions; and (iii) thermal time constants, and their effect on phase wander and on the Allan variance. 

\vspace{1em}\textbf{Keywords:}
Phase Noise, Jitter, Aliasing, FPGA, Bouncing, Allan Variance, Thermal Stability.

\end{abstract}

\section{Introduction}\label{sec:Introduction}
Timing analysis is generally driven by the design of logic functions.  That is why specs like ``the input must be stable 600 ps before the clock edge'' are just countless.  
From this standpoint, it is sufficient to describe the fluctuations in terms of \emph{jitter}.
Broadly speaking, jitter is the time fluctuation, evaluated in reference conditions. 
Because of the wide bandwidth, jitter is chiefly determined by the white noise.  Notice that proper operation requires an analog bandwidth 3--4 times the switching frequency, and in turn up to a few GHz with nowadays components. 

When the design comes to spectral analysis and to highly stable oscillators, language and requirements change radically. 
Fluctuations are generally described in terms of phase noise, expressed either as $S_\varphi(f)$ or $\mathscr{L}(f)$, and the low-frequency phenomena are no longer negligible.
Low phase noise is crucial in radars \cite{Skolnik-2002,Skolnik-2008,Krieger-2006}, modern telecomm \cite{Esman-2016}, atomic frequency standards \cite{FSM-2015} and particle accelerators \cite{Serrano-2011,Jablonski-2015}, just to mention some.

In the rapidly changing world of digital electronics, the literature on phase noise is rather old and focuses on frequency dividers, either in TTL and ECL components \cite{Phillips-1987,Egan-1990}, or in transistor-level modeling.  Other references found are more about data transfer in telecom networks than about components \cite{Reinhardt-2005,Bregni-2002,Kihara-1989}.

At the time of \cite{Phillips-1987,Egan-1990}, CMOS technology was used only in microprocessors and complex functions. 
Gate arrays and FPGAs came later, with a new rapid progress
\cite{Mack-2015,Huang-2015,Moore-50y}.  
Interestingly for us, gate arrays and FPGAs bridge the gap between logical/computational functions and circuit-level design.  The precise control on electrical signals that follows opens a new challenge in understanding noise.
However, VLSI engineers are mostly concerned with noise margin, crosstalk, and power distribution \cite{Weste-2011}.  Conversely, amplitude and phase noise are not studied.

The purpose of this article is to set the basic knowledge about phase noise, and to provide examples.  We focus on the clock distribution because clock edges are the most critical ones for timing.  This does not sounds a limitation, first because critical signals can be synchronized to a clock line, and second because a chip in charge of a highly critical operation should not perform multiple tasks `cross-talking' at random with one another.

Designing the experiments was initially difficult.  However, after a noise model and the first results were available, reproducing similar experiments is surprisingly simple.  We hope that the reader will be able to port our ideas to other technologies and logic families.  The reader may also learn about reverse engineering the noise.

\section{Definitions, and Phase Noise Models}\label{sec:Noise-model}
\begin{sidewaystable}\centering\small\sffamily
\caption{Phase Noise Types and Their Parameters}\label{tab:noise-types}
\newcommand{\vra}{\rule[-2ex]{0ex}{5ex}}
\newcommand{\vrb}{\rule[-1.5ex]{0ex}{4ex}}
\newcommand{\vrc}{\rule[-3.5ex]{0ex}{8ex}}
\newcommand{\vrd}{\rule[-5.ex]{0ex}{11.5ex}}
\newcommand{\vre}{\rule[-2.5ex]{0ex}{6ex}}
\newcommand{\vrf}{\rule[-2.5ex]{0ex}{7ex}}
\begin{tabular}{|c|c|c|cc|cc|cc|}\hline
\vra Noise Type 
&\multicolumn{2}{c|}{Dependence on $\nu_0$}
&\multicolumn{2}{c|}{Main Equation} 
&\multicolumn{2}{c|}{Derived Equation} 
&\multicolumn{2}{c|}{\parbox{9em}{\centering Parameters}} \\\cline{2-3}
&\vrb$S_\varphi(f)$	&$S_\mathsf{x}(f)$&&&&&&\\\hline
\vrc\parbox{9em}{\centering\sffamily Pure phase type\\[0.5ex](pure $\varphi$-type)}
&$C$&$1/\nu_0^2$
&$\mathsf{b}_{-1}=\dfrac{\mathrm{h}_{-1}}{V_0^2}$&\req{eqn:pure-phi-type}
&$\mathsf{k}_{-1}=\dfrac{\mathrm{h}_{-1}}{4\pi^2\nu_0^2V_0^2}$&\req{eqn:pure-phi-type-x}
&~~\raisebox{2ex}{\makebox[0pt]{$\sqrt{\mathrm{h}_{-1}}$}}%
\raisebox{-2ex}{\makebox[0pt]{$V_{0}$}}
&~~\raisebox{2ex}{\makebox[0pt]{\unit{[V]}}}%
\raisebox{-2ex}{\makebox[0pt]{\unit{[V]}}}
\\\hline
\vrc \parbox{9em}{\centering Aliased phase type\\[0.5ex](aliased $\varphi$-type)}
&$1/\nu_0$  & $1/\nu_0^3$
&$\mathsf{b}_0=\dfrac{B\,\mathrm{h}_{0}}{\nu_0V_0^2}$&\req{eqn:aliased-phi-type}
&$\mathsf{k}_0=\dfrac{\mathrm{h}_{0}\,B}{4\pi^2\nu_0^3V_0^2}$&\req{eqn:aliased-phi-type-x}
&\raisebox{2ex}{\makebox[0pt]{$\sqrt{\mathrm{h}_{0}B}$}}%
\raisebox{-2ex}{\makebox[0pt]{$V_{0}$}}%
&\raisebox{2ex}{~~\makebox[0pt]{\unit{[V]}}}%
\raisebox{-2ex}{\makebox[0pt]{\unit{[V]}}}%
\\\hline
\vre \parbox{9em}{\centering Pure time type\\[0.5ex](pure $\mathsf{x}$-type)}
& $\nu_0^2$	&	$C$
&$\mathsf{k}_{-1}=C$&\req{eqn:pure-x-type}
&$\mathsf{b}_{-1}=4\pi^2\nu_0^2\,\mathsf{k}_{-1}$&\req{eqn:pure-x-type-phi}
&$\sqrt{\mathsf{k}_{-1}}$&\unit{[s]}\\\hline
\vrf \parbox{9em}{\centering Aliased time type\\[0.5ex](aliased $\mathsf{x}$-type)}
& $\nu_0$&$1/\nu_0$
&$\mathsf{k}_{0}=\dfrac{\mathsf{J}^2}{\nu_0}$&\req{eqn:aliased-x-type}
&$\mathsf{b}_0=4\pi^2\mathsf{J}^2\nu_0$&\req{eqn:aliased-x-type-phi}
&$\mathsf{J}$&\unit{[s]}\\\hline
\end{tabular}
\end{sidewaystable}

Phase noise is often expressed as the one-sided PSD $S_\varphi(f)$ of the random phase $\varphi(t)$.
In technical literature we often find $\mathscr{L}(f)$, defined as $\mathscr{L}(f)=\frac{1}{2}S_\varphi(f)$ and given in dBc/Hz \cite{IEEE-STD-1139-2008}. Alternatively, phase noise is represented as the phase time fluctuation $\mathsf{x}(t)$, and its PSD $S_\mathsf{x}(f)$.
Since $\mathsf{x}(t)$ is equivalent to $\varphi(t)$ converted into time, it holds that 
\begin{align}
\label{eqn:x}
\mathsf{x}(t)&=\frac{1}{2\pi\nu_0}\varphi(t)&&\unit{[s]}\\[1ex]
\label{eqn:Sx}
S_\mathsf{x}(f)&=\frac{1}{4\pi^2\nu_0^2}S_\varphi(f)&&\unit{[s^2\!/Hz]}\,,
\end{align}
where $\nu_0$ is the carrier frequency.  Our notation is consistent with general literature \cite{IEEE-STD-1139-2008,ccir90rep580-3}, yet for the choice of fonts for some specific quantities as a minor detail.

A model which is useful to describe phase noise is the polynomial law
\begin{align}
S_\varphi(f)&=\sum_{j=m}^{0}\;\mathsf{b}_jf^j &
S_\mathsf{x}(f)
&=\sum_{j=m}^{0}\;\mathsf{k}_jf^j\,,
\label{eqn:poly}
\end{align}
where the integer $m<0$ depends on the device.  
After \req{eqn:Sx}, it holds that $\mathsf{k}_j=\mathsf{b}_j/4\pi^2\nu_0^2$.
The sum \req{eqn:poly} describes the usual noise types: white phase noise $\mathsf{b}_0$, flicker phase noise $\mathsf{b}_{-1}/f$, white frequency noise $\mathsf{b}_{-2}/f^2$, etc.
Common sense suggests that in two-port components, noise processes higher than $1/f$ (i.e., $f^j$,  $j{<}-1$) cannot extend over unlimitedly low frequencies, otherwise  the input-output delay diverges in the long run.

The polynomial law is also used for the PSD of the voltage noise $n(t)$
\begin{align}
S_n(f)=\sum_{j=m}^{0}\;\mathrm{h}_jf^j\qquad\unit{[V^2\!/Hz]}
\end{align}
(notice the font in $\mathrm{h}_j$, because $\mathsf{h}_j$ reserved for $S_\mathsf{y}(f)=\sum_{j}\mathsf{h}_jf^j$). 
The reader familiar with analog electronics finds an obvious analogy with the parameter $e_n$ [\unit{nV/\sqrt{Hz}}], specified separately for white and flicker noise.

The rms time fluctuation $\mathsf{J}$ can be calculated integrating $S_\mathsf{x}(f)$ over the system bandwidth (Parseval theorem)
\begin{equation}
\mathsf{J}^2=\int_{f_L}^{f_H}S_\mathsf{x}(f)\:df\,.
\label{eqn:fluctuation-J} 
\end{equation}
The lower limit $f_L$ is set by maximum differential delay in the system. 
The upper limit is $f_H=\nu_0$.  The reason is that the fluctuations are sampled at the clock edges, thus at $2\nu_0$.
The quantity $\mathsf{J}^2$ can be identified with the variance $\left<\mathsf{x}^2(t)\right>$, yet after filtering out the $f<f_L$ part. 

For our purposes, $\mathsf{J}$ is approximately equivalent to the \emph{rms jitter}.  By contrast, the general term `jitter' has wider scope, mostly oriented to SDH telecomm systems.  It includes different types of noise and interferences starting at 10 Hz, with different weight for each (the term `wander' is preferred below 10 Hz). See for example \cite{G.8252,Li-2008,Reinhardt-2005} for standards and useful digressions.  In a FPGA, there may be a factor 1000 between the rms jitter and the overall jitter, also including interferences.

We introduce two basic types of process discussed below, which take their names from the frequency-scaling properties.

The \emph{\bfseries phase-type} (or \emph{pure phase-type}) process is, \emph{by definition}, a process in which the statistical properties of $\varphi(t)$ are unaffected after changing the carrier frequency $\nu_0$ in a suitable wide range.  
Hence, $\mathsf{x}(t)$ scales with $\nu_0$ according to \req{eqn:x}.  

The roles of $\varphi(t)$ and $\mathsf{x}(t)$ are interchanged in the time-type process.  
So, the \emph{\bfseries time-type} (or \emph{pure time-type}) process is, \emph{by definition}, a process in which the statistical properties of $\mathsf{x}(t)$ are unaffected after changing the carrier frequency $\nu_0$ in a suitable wide range.  
Of course, $\mathsf{x}(t)$ scales according to \req{eqn:x}.

The concepts of phase-type and time-type process apply to phase noise, wavelet variances (Allan and Allan-like), environmental effects, etc.
Most readers are familiar with the `personality' of the $\varphi$-type noise from the phase noise of RF/microwave amplifiers \cite{Boudot-2012}.  
Thermal noise, flicker, and some environmental effects in amplifiers behave in this way.  
Conversely, the thermal drift of the delay in a coaxial cable or optical fiber are time-type processes.
The $\mathsf{x}$-type noise also describes the ideal noise-free synthesizer, which transfers $\mathsf{x}(t)$ from the input to the output, independently of $\nu_0$.

\section{Noise in the Clock Distribution}\label{ssec:processes}
\begin{figure}
\centering\includegraphics[width=0.64\columnwidth]{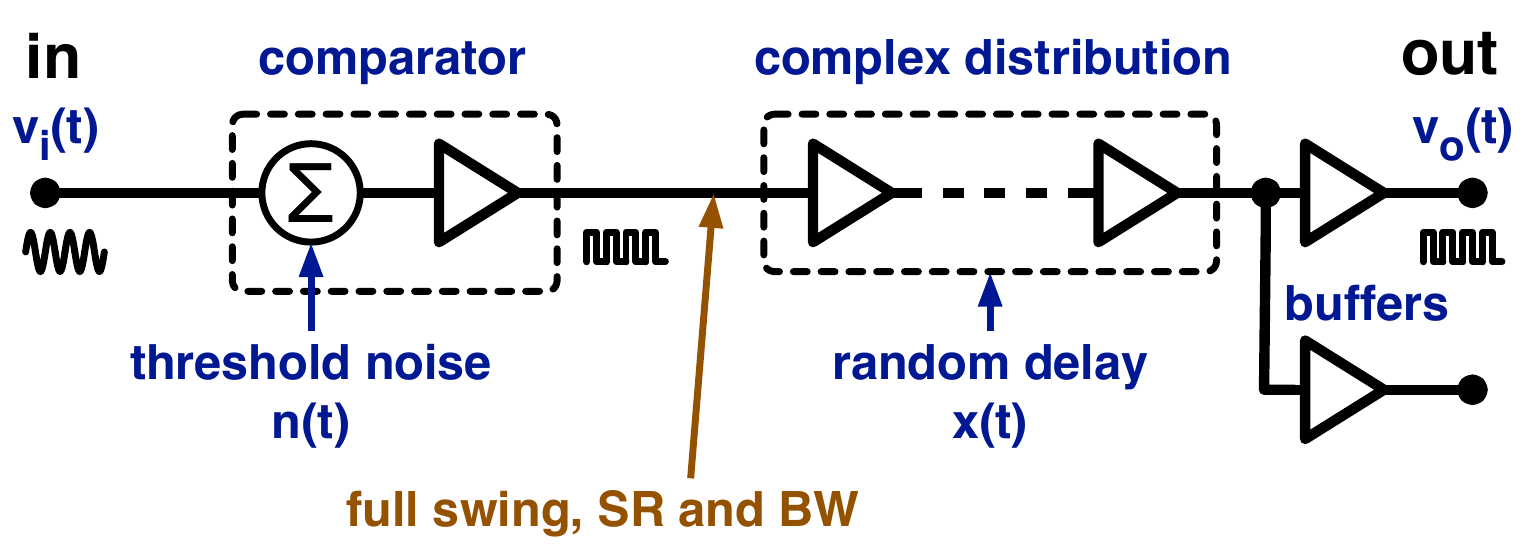}
\caption{Block diagram describing the noise in the clock distribution.}
\label{fig:Clock-digram}
\end{figure}

A lot about phase and time fluctuations can be learned from the simple model sketched in Fig.\,\ref{fig:Clock-digram}.
The input signal of frequency $\nu_0$ is first converted into a square wave with full voltage swing, full slew rate and full bandwidth, and then distributed.  
Restricting our attention to white and flicker, we get the four behaviors listed in Table~\ref{tab:noise-types} and discussed below.

\subsection{Spectrum of the Phase-Type ($\varphi$-type) Phase Noise}\label{ssec:phi-type}
In digital circuits we often encounter the \emph{aliased} $\varphi$-type noise.
Let us start with $\varphi$-type noise at the input of a digital circuit, where the input signal $v(t)$ crosses a threshold affected by a fluctuation $n(t)$.  Under the assumption that the input Slew Rate (SR) is high enough to avoid multiple bouncing (Sec.\,\ref{sec:Chatter}), we get
$\mathsf{x}(t)=n(t)/\mathrm{SR}$ and, after \req{eqn:x},
\begin{align}
\varphi(t)&=\frac{2\pi\nu_0}{\mathrm{SR}}\,n(t)\,.
\label{eqn:n-to-phi-conv}
\end{align}
Notice that the direct measurement of $n(t)$ is possible only in simple circuits which allow the simultaneous access to input and output of the gate.  

The sinusoid is the preferred clock waveform because it propagates through circuit boards with best impedance matching and lowest crosstalk and radiation, and because high purity reference oscillators work in sinusoidal regime.
Discarding the dc component and setting the threshold at 0, the clock signal
\begin{align}
v(t)=V_0\cos(2\pi\nu_0t)
\label{eqn:clock}
\end{align}
has slew rate $\mathrm{SR}_v=2\pi\nu_0V_0$.
In this conditions, the phase fluctuation is
\begin{align}
\label{eqn:phi-type}
\varphi(t)&=\frac{n(t)}{V_0}
&\text{($\varphi$-type)}\,.
\end{align}

Generally, the analog bandwidth $B$ of a digital circuit is greater than the max $\nu_{0}$ by a factor of 3--4.  This is necessary for the device to switch correctly.  In turn, the bandwidth of $n(t)$ is equal to $B$.
Squaring the input signal samples $n(t)$ at the zero crossings introduces aliasing.  The spectrum of the sampled signal is
\begin{align}
S_{n,s}(f)&=\frac{B}{\nu_0}\,\mathrm{h}_{0}+\ldots
&&\text{(sampled noise)}\,,
\label{eqn:aliasing}
\end{align}
where the $1/f$ and higher terms are neglected because of the comparatively noise power.
A trivial way to prove \req{eqn:aliasing} is to calculate the variance 
$\left<n^2(t)\right>=\mathrm{h_0}B$ (Parseval theorem) before sampling, and to state that it is equal to the variance $\sigma^2=S_{n,s}(f)\nu_0$ of the sampled signal.
Accordingly, the phase noise is
\begin{align}
\label{eqn:aliased-phi-type}
\mathsf{b}_0&=\frac{\mathrm{h}_{0}\,B}{\nu_0V_0^2}
&&\text{(white, aliased $\varphi$-type)}\\[1ex]
\label{eqn:aliased-phi-type-x}
\mathsf{k}_0&=\frac{\mathrm{h}_{0}\,B}{4\pi^2\nu_0^3V_0^2}
&&\text{(same, after \req{eqn:Sx})}\,.
\end{align}

Oppositely, aliasing has negligible effect on flicker $\mathrm{h}_{-1}/f$ and on higher terms ($1/f^2$, $1/f^3$ etc.).  It follows from \req{eqn:phi-type} that
\begin{align}
\label{eqn:pure-phi-type}
\mathsf{b}_{-1}&=\frac{\mathrm{h}_{-1}}{V_0^2}\,,
\qquad C~\text{vs.}~\nu_0
&&\text{(flicker, pure $\varphi$-type)}\\[1ex]
\label{eqn:pure-phi-type-x}
\mathsf{k}_{-1}&=\frac{\mathrm{h}_{-1}}{4\pi^2\nu_0^2V_0^2}
&&\text{(same, after \req{eqn:Sx})}\,.
\end{align}

\begin{figure}[t]
\centering\includegraphics[width=0.64\columnwidth]{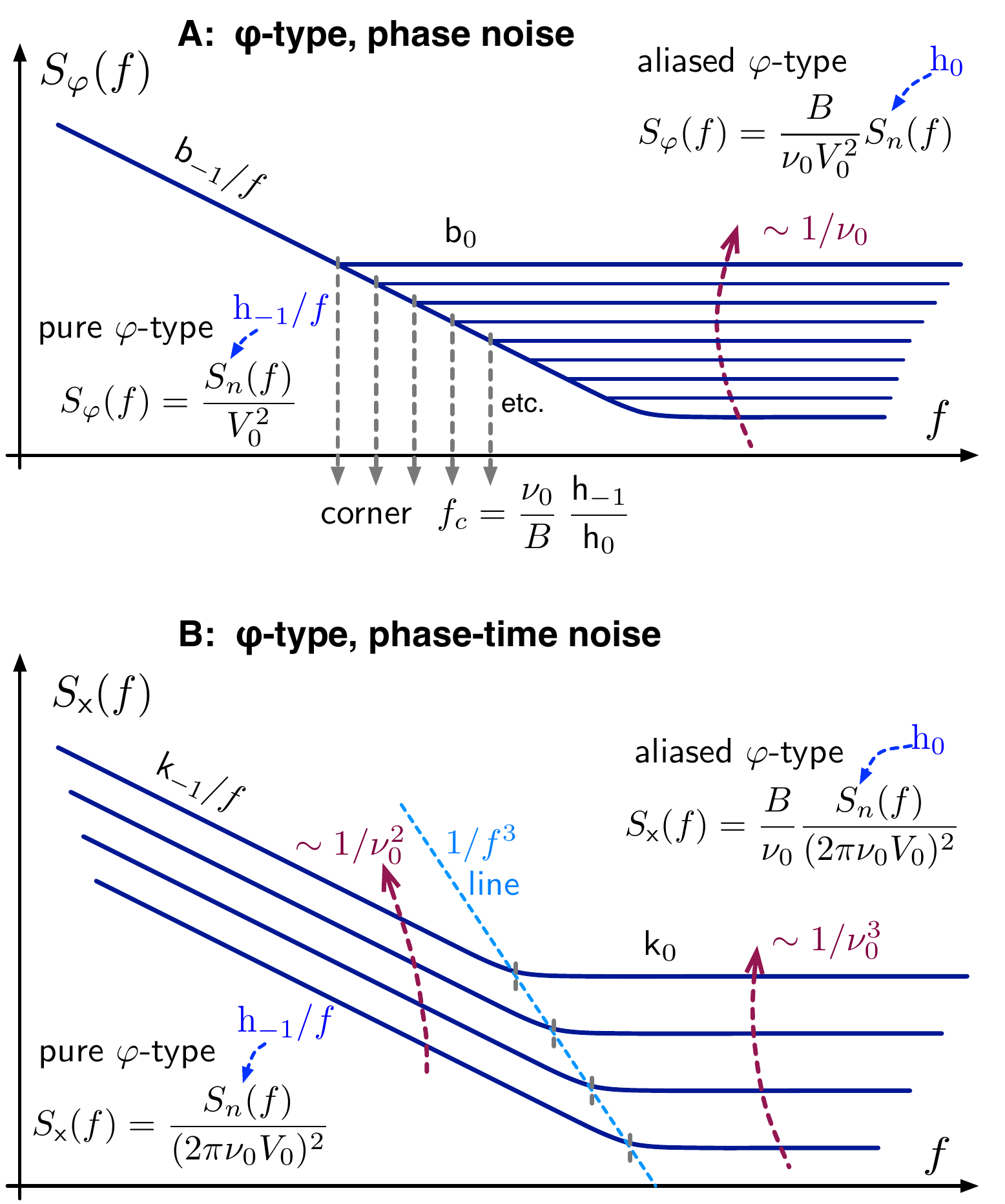}
\caption{Spectra originated by the phase type ($\varphi$-type) phase noise.}
\label{fig:phi-type}
\end{figure}
Figure~\ref{fig:phi-type} shows the spectral properties of the $\varphi$-type noise.  
Aliasing scales the white noise as $1/\nu_0$, but it has no effect on flicker.  The corner frequency $f_c$ which separates white from flicker regions is obtained equating \req{eqn:aliased-phi-type} to \req{eqn:pure-phi-type}
\begin{align}
\label{eqn:corner-phi}
f_c&=\frac{\nu_0}{B}\:\frac{\mathrm{h}_{-1}}{\mathrm{h}_0}
\qquad\text{(corner, $\varphi$-type noise)}\,.
\end{align}

\subsection{Spectrum of the Time Type ($\mathsf{x}$-type) Phase Noise}\label{ssec:x-type}
The $\mathsf{x}$-type noise originates after the input comparator, where the clock signal has full SR and bandwidth.
Though threshold fluctuations are always present, the voltage-to-time conversion has little effect, and the gate is characterized by its delay fluctuations.
So, each gate of the clock distribution contributes to the delay, and the fluctuations add up statistically.
At a closer sight, the device may be organized hierarchically, for example in gates and cells, likely with a longer propagation time between cells.
Nonetheless, the fluctuation is proportional to the length and to the complexity of the distribution chain. 

The pure $\mathsf{x}$-type noise is found in the $1/f$ region and below, not affected by aliasing.  The noise spectrum is described by
\begin{align}
\label{eqn:pure-x-type}
\mathsf{k}_{-1}&=C~\text{vs.\ $\nu_0$}
&&\text{(flicker, pure \textsf{x}-type)}\\[1ex]
\label{eqn:pure-x-type-phi}
\mathsf{b}_{-1}&=4\pi^2\nu_0^2\,\mathsf{k}_{-1}
&&\text{(same, after \req{eqn:Sx})}\,,
\end{align}
where $\mathsf{k}_{-1}$ is the technical parameter which results from the clock distribution.

The aliased $\mathsf{x}$-type results from sampling the fluctuation at the frequency $2\nu_0$, which affects the white noise region.  The spectral parameter $\mathsf{k}_0$ is found in the same way as with \req{eqn:aliasing}, neglecting the $1/f$ and higher terms
\begin{align}
\label{eqn:aliased-x-type}
\mathsf{k}_0&=\mathsf{J}^2/\nu_0
&&\text{(white, aliased \textsf{x}-type)}\\[1ex]
\label{eqn:aliased-x-type-phi}
\mathsf{b}_0&=4\pi^2\mathsf{J}^2\nu_0
&&\text{(same, after \req{eqn:Sx})}\,.
\end{align}

\begin{figure}
\centering\includegraphics[width=0.64\columnwidth]{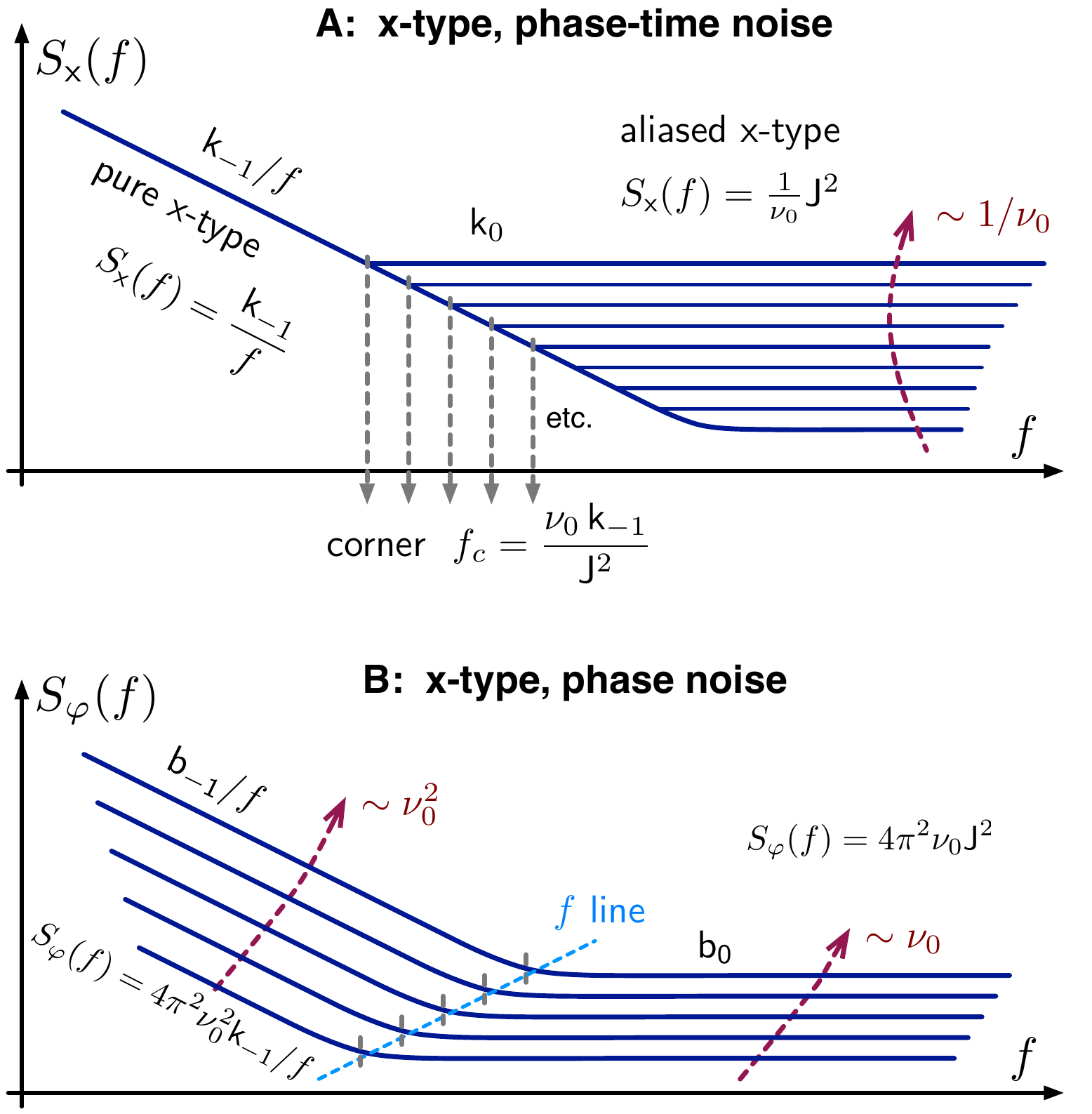}
\caption{Spectra originated by the time type ($\mathsf{x}$-type) phase noise.}
\label{fig:x-type}
\end{figure}
The spectral properties of the $\mathsf{x}$-type noise --- i.e., \req{eqn:pure-x-type}--\req{eqn:aliased-x-type-phi} --- are summarized in Fig.\,\ref{fig:x-type}.
The corner frequency which divides the flicker from the white region is calculated by equating \req{eqn:pure-x-type} to \req{eqn:aliased-x-type} 
\begin{align}
\label{eqn:corner-x}
f_c&=\frac{\nu_0\,\mathsf{k}_{-1}}{\mathsf{J}^2}
\qquad\text{(corner, $\mathsf{x}$-type noise)}\,.
\end{align}

\subsection{Interpretation of Phase Noise Spectra}
\begin{figure}
\centering\includegraphics[width=0.71\columnwidth]{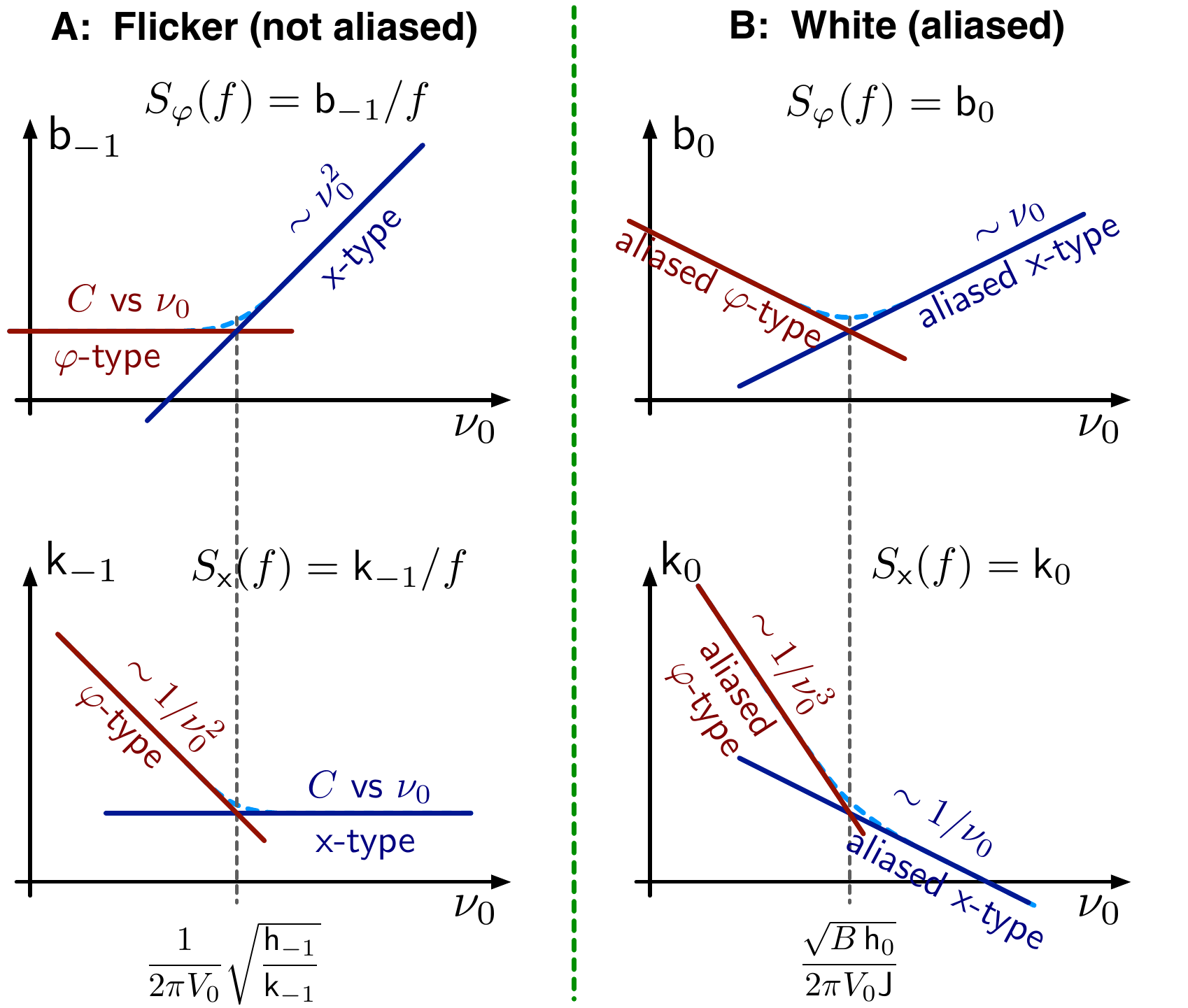}
\caption{Comparison between $\varphi$-type and $\mathsf{x}$-type noise.}
\label{fig:phi-vs-x-type}
\end{figure}
A series of spectra $S_\varphi(f)$ taken with several values of $\nu_0$ helps to understand the interplay of noise types.
Scaling $\nu_0$ in powers of two seems appropriate.

Let us start with flicker, $S_\varphi(f)=\mathsf{b}_{-1}/f$.  Comparing \req{eqn:pure-phi-type} to \req{eqn:pure-x-type-phi}, we expect that the noise is of the $\varphi$-type at low $\nu_0$, and of the $\mathsf{x}$-type at high $\nu_0$, with a corner frequency 
\begin{align}
\nu_c&=\frac{1}{2\pi V_0}\sqrt{\frac{\mathrm{h}_{-1}}{\mathsf{k}_{-1}}}
&&\text{(flicker)}\,.
\label{eqn:thresh-flicker}
\end{align}
This is shown in Fig.\,\ref{fig:phi-vs-x-type}\,A\@.
Far from $\nu_c$, we can evaluate
\begin{align}
\mathrm{h}_{-1}&=V_0^2\,\mathsf{b}_{-1}
&&\text{($\nu_0\ll\nu_c$)}\\[1ex]
\mathsf{k}_{-1}&=\frac{\mathsf{b}_{-1}}{4\pi^2\nu_0^2}
&&\text{($\nu_0\gg\nu_c$)}\,.
\end{align}

The white phase noise $S_\varphi(f)=\mathsf{b}_0$ is described by \req{eqn:aliased-phi-type} at low $\nu_0$, and by \req{eqn:aliased-x-type-phi} at high $\nu_0$, separated by the cutoff
\begin{align}
\nu_c&=\frac{\sqrt{B\,\mathrm{h}_0}}{2\pi V_0\mathsf{J}}.
&&\text{(white)}\,.
\label{eqn:thresh-white}
\end{align}
This is shown on Fig.\,\ref{fig:phi-vs-x-type}\,B\@.
At low $\nu_0$, \req{eqn:aliased-phi-type} enables to calculate the noise power $\left<n^2(t)\right>=\mathrm{h}_{0}\,B$ of the input threshold
\begin{align}
\mathrm{h}_{0}\,B&=V_0^2\,\mathsf{b}_0\nu_0
&&\text{($\nu_0\ll\nu_c$)}\,.
\end{align}
Assuming that $B$ is equal to 3--4 times the maximum $\nu_0$, we can infer $\mathrm{h}_{0}$ and the noise voltage $e_n=\sqrt{\mathrm{h}_{0}}$.
Conversely, at high $\nu_0$ we can extract the fluctuation
\begin{align}
\mathsf{J}&=\frac{1}{2\pi}\sqrt{\frac{\mathsf{b}_0}{\nu_0}}
&&\text{($\nu_0\gg\nu_c$)}\,.
\end{align}
This can be compared to the rms jitter, if available in the specs.

\section{Selected Noise Measurements}\label{sec:Selected-devices}
\begin{figure}
\centering\includegraphics[width=0.64\columnwidth]{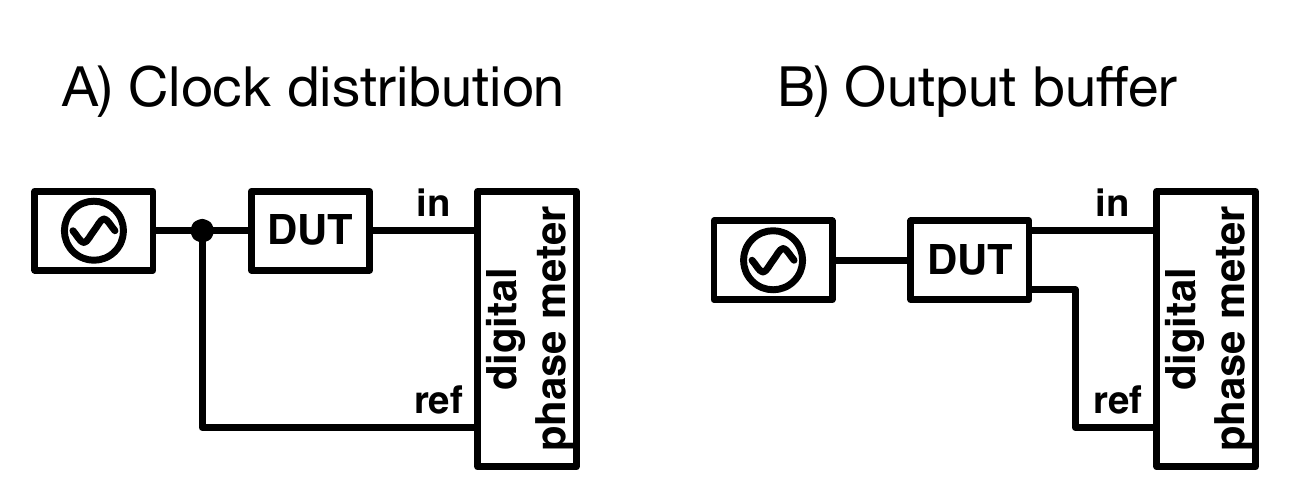}
\caption{The digital phase meter is either a Symmetricom (now Microsemi) 5125 or 5120.  The two outputs may have different frequency.}
\label{fig:FPGA-Method}
\end{figure}
We measured the phase noise of several devices routinely used in our labs.  This is a necessary step, before considering an unbound search for the best.
Accordingly, the measurement method (Fig.\,\ref{fig:FPGA-Method}) is more about flexibility than about sensitivity.  Anyway, the phase noise of digital components is generally higher than that of common low noise components (i.e., amplifiers and mixers).  On the other hand, we need simple operation in a wide range of frequency, with signals that may \emph{not} be at the same frequency as the reference.  For us, this is the relevant feature of the Microsemi 5125 (1--400 MHz) and 5120 (1--30 MHz) instruments.  These instruments make use of correlation and average on the spectra of two nominally equal channels which measure the same quantity, which rejects the single channel noise \cite{Rubiola-2010:xsp,Stein-2010:avar}.  Notice that the oscillator is common mode, with very small differential delay, hence its noise is highly rejected.
The Fourier frequency spans from 1 mHz to 1 MHz.

\subsection{Cyclone III (65 nm)}\label{Cyclone-III}
\begin{figure}[t]
\centering
\includegraphics[width=0.84\columnwidth]{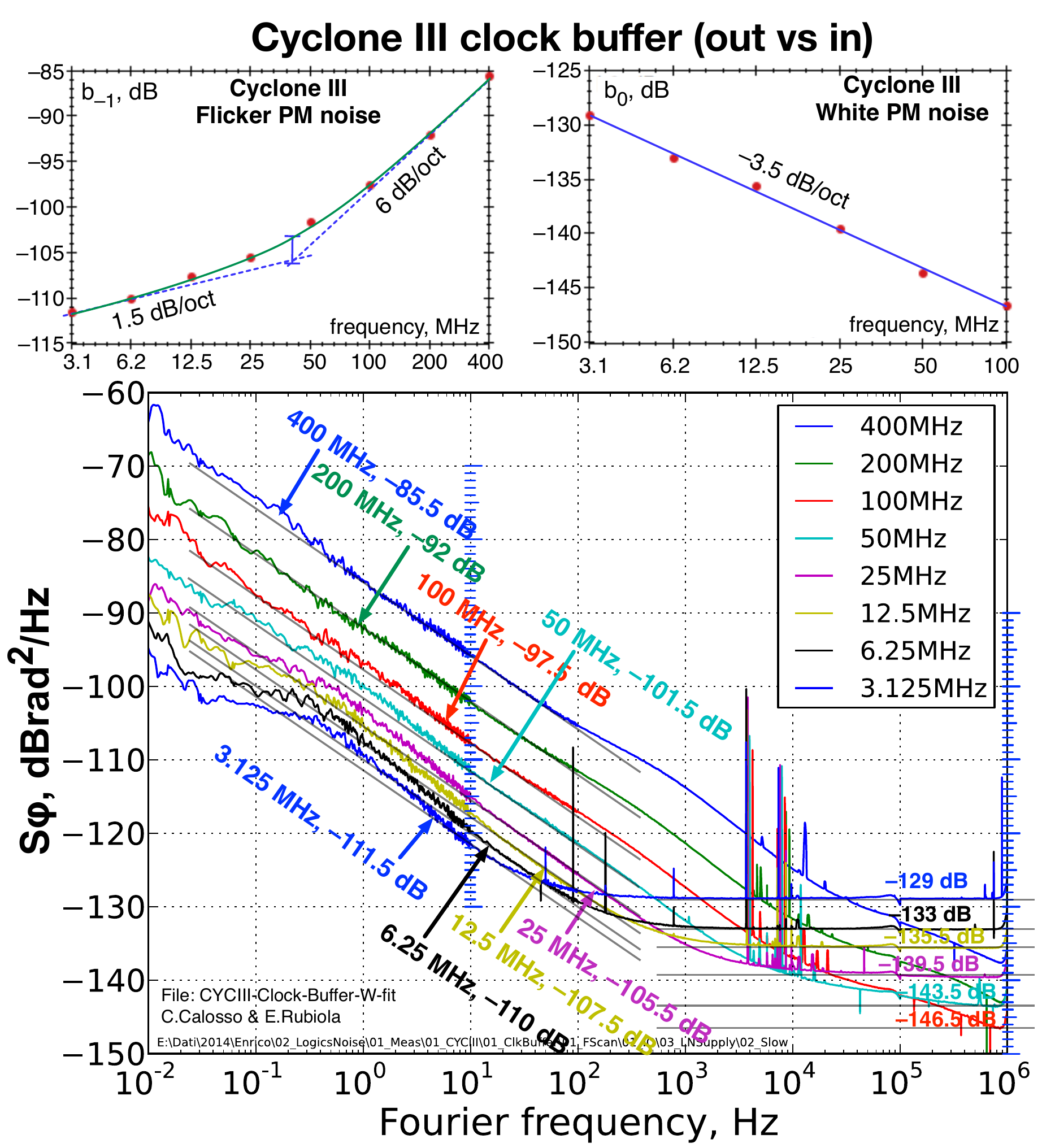}
\caption{Phase noise of the Cyclone III clock distribution.}
\label{fig:CYCIII-Clock-buffer-W-fit}
\end{figure}

In a first experiment, we measure a Cyclone III \cite{Cyclone-III} in a clock buffer configuration. The input sinusoidal clock $V_0=1$ \unit{V_\text{\!peak}} ($+10$ dBm on 50 \ohm) is squared and distributed as in Fig.\,\ref{fig:Clock-digram}\,A\@.  The spectrum is shown in Fig.\,\ref{fig:CYCIII-Clock-buffer-W-fit}.

We first look at the white noise region.  Our model suggests aliased $\varphi$-type noise \req{eqn:aliased-phi-type} at low $\nu_0$, and aliased $\mathsf{x}$-type noise \req{eqn:aliased-x-type-phi} beyond the cutoff given by \req{eqn:thresh-white}, as shown on Fig.\,\ref{fig:phi-vs-x-type}\,B\@.
Starting from $\nu_0=3.125$ MHz, $\mathsf{b}_0$ scales down as $-3.5$ dB per factor-of-two, in fairly good agreement with the 3 dB predicted by the model.  This results from  
the data fit shown on Fig.\,\ref{fig:CYCIII-Clock-buffer-W-fit} top-right.
Taking $V_0=1$ V, \req{eqn:aliased-phi-type} gives a threshold fluctuation $\sqrt{\mathrm{h}_0B}=550\pm65$ \unit{\mu V}.  The `$\pm\,65$ \unit{\mu V}' results from $\mathsf{b}_0\propto1/\nu_0^{1.16}$' instead of the $1/\nu_0$ law.
Assuming $B=2.5$ GHz (analog bandwidth, four times the maximum toggling frequency), we get $\sqrt{\mathrm{h}_0}=11\pm1.3$ \unit{nV/\sqrt{Hz}}.  This is in agreement with general experience, which suggests that general high-speed electronics has a typical noise level of 10--15 \unit{nV/\sqrt{Hz}}.

At $\nu_0\ge100$ MHz, the white noise falls outside the 1 MHz span.  Since this occultation occurs before the aliased $\mathsf{x}$-type noise shows up, we have no direct access to $\mathsf{k}_0$.  
On Fig.\,\ref{fig:CYCIII-Clock-buffer-W-fit}, at the maximum $f$ (1 MHz) and at 400 MHz carrier, the white noise is below $-138$ \unit{dBrad^2/Hz} (upper bound).  This value, integrated over $B=400$ MHz and converted into time, gives 1 ps, which is an upper bound for $\mathsf{J}$.


Flicker noise is in good agreement with pattern of Fig.\,\ref{fig:phi-vs-x-type}\,B only at $\nu_0\ge100$ MHz.  From this part of the plot, we calculate $\sqrt{\textsf{k}_{-1}}=21$ fs.
By contrast, at $\nu_0\le50$ MHz $\mathsf{b}_{-1}$ scales as $\approx1.5$ dB per factor-of-two instead of being constant.  This discrepancy is not understood.  However, the $1/f$ region is rather irregular, and corrupted by bumps, even more pronounced at low $\nu_0$.

The lowest flicker found on Fig.\,\ref{fig:CYCIII-Clock-buffer-W-fit} ($-115$ \unit{dBrad^2/Hz} at 3.125 MHz carrier), converted into voltage using \req{eqn:pure-phi-type}, gives $\sqrt{\mathrm{h}_{-1}}=2.6$ \unit{\mu V} (upper bound for the input voltage flicker).  Interestingly, this value is similar to the flicker of some CMOS high-speed operational amplifiers (for instance, 1.9 \unit{\mu V} for the Texas Instruments OPA354A).

\begin{figure}[t]
\centering
\includegraphics[width=0.84\columnwidth]{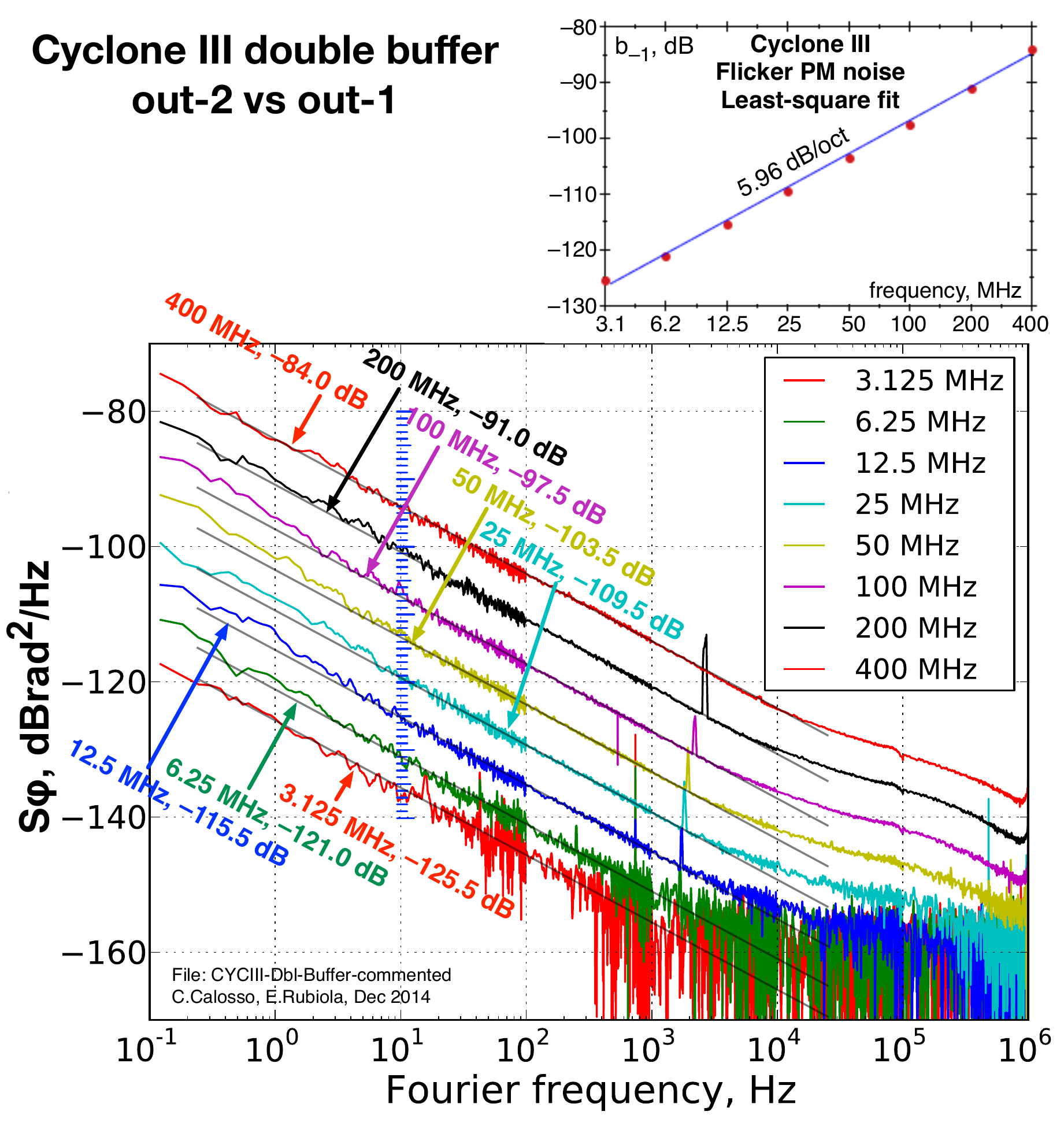}
\caption{Phase noise of the Cyclone III, measured by comparing two outputs.  Take away 3 dB for the noise of one buffer.}
\label{fig:CYCIII-Dbl-Buffer-commented}
\end{figure}

Figure\,\ref{fig:CYCIII-Dbl-Buffer-commented} shows the phase noise of the output buffer.  The white noise is too low to be visible with the 1 MHz span, masked by flicker and by some bumps at $10^4\ldots10^6$ Hz.
By contrast, the flicker noise is in perfect agreement with the 6 dB per factor-of-two model (pure $\mathsf{x}$-type noise).
Comparing 
Fig.\,\ref{fig:CYCIII-Dbl-Buffer-commented} to 
Fig\,\ref{fig:CYCIII-Clock-buffer-W-fit}, at $\nu_0=400$ MHz the flicker of the complete clock distribution is close to that of the output buffer.  So, the contribution of the output buffer is not negligible.
Conversely, at lower $\nu_0$ a significantly larger flicker rises in the clock distribution chain.

\subsection{Measuring the Time Type ($\mathsf{x}$-Type) Noise with the $\Lambda$ Divider}
After some tests, we realized that the $\Lambda$ frequency divider \cite{Calosso-2013} is a good tool to measure the $\mathsf{x}$-type noise of the clock distribution.  First, a frequency divider is useful in that the input time fluctuation ($\varphi$-type noise, \req{eqn:pure-phi-type-x}) is kept low by using a high input frequency, while the measurement at the lower output frequency is simpler (both instruments are suitable, and the background is lower).
Second, the $\Lambda$ divider circumvents the aliasing phenomenon.  In fact, a $\Lambda$ divider $\div\mathcal{D}$ provides a triangle-like output waveform by combining $\mathcal{D}$ phases of a square wave, which is equivalent to sampling at the input frequency. 
\begin{figure}
\centering
\includegraphics[width=0.84\columnwidth]{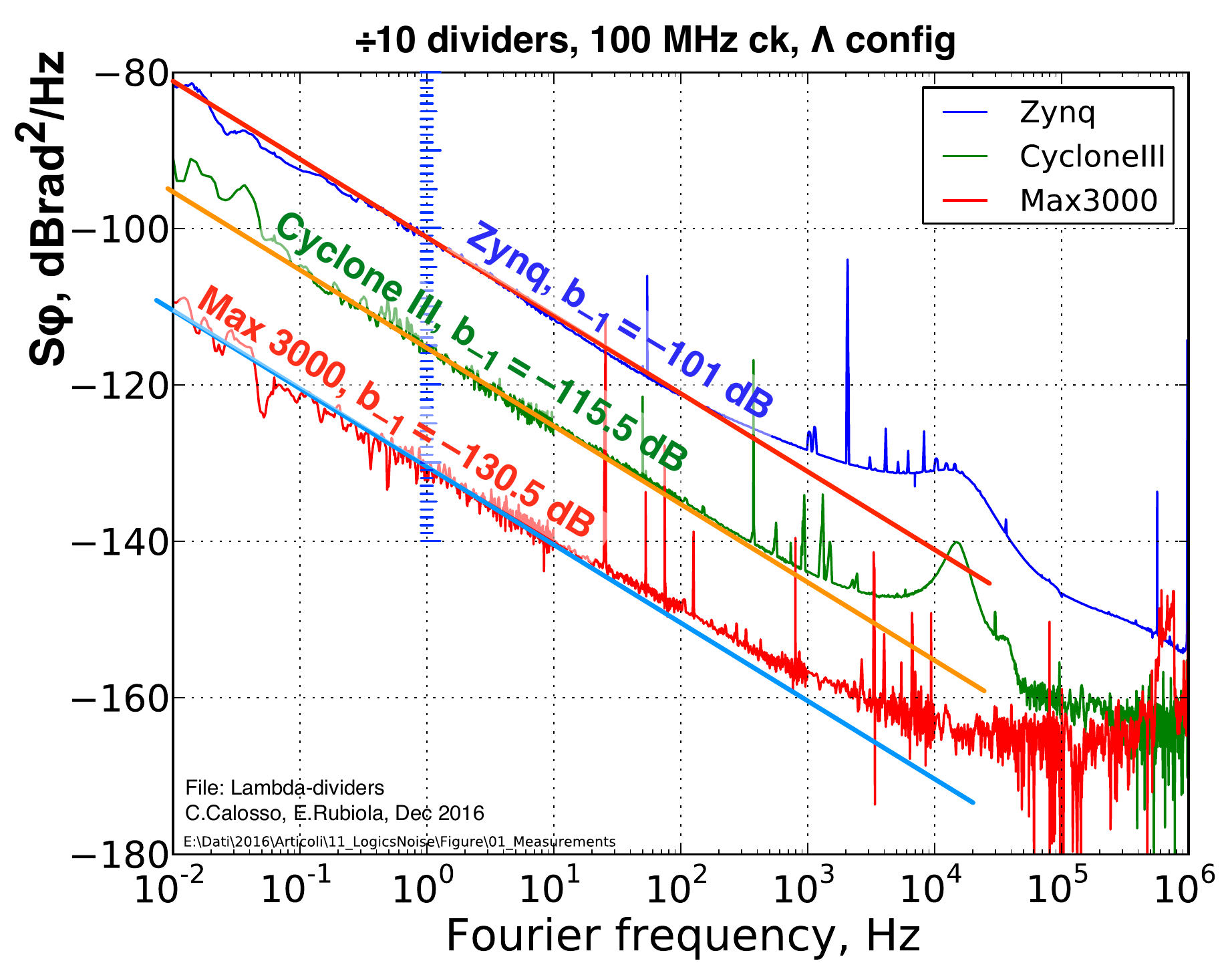}
\caption{Phase noise of some components used as a $\div10$ frequency divider in the $\Lambda$ configuration.}
\label{fig:Lambda-dividers}
\end{figure}

Figure \ref{fig:Lambda-dividers} shows the phase noise of some devices used as $\div10$ dividers in $\Lambda$ configuration, with 100 MHz input and 10 MHz output frequency.  The flicker coefficient is clearly identified, not corrupted by artifacts.  The bump at 20 kHz (Zynq and Cyclone III) is due to the insufficiently filtered power supply.  Finally, the $\Lambda$ divider implemented with the Max 3000 deserves mentioning for its low noise ($\mathsf{b}_{-1}=-130.5$ \unit{dBrad} and $\mathsf{b}_{0}=-165$ \unit{dBrad^2/Hz}).  This is lower than regular dividers (general experience), and just 10 dB above the NIST regenerative dividers \cite{Hati-2013} at the same output frequency.


\section{The Volume Law}\label{sec:Volume-law}
The idea that the phase noise coefficient $\mathsf{b}_{-1}$ is proportional to $1/\mathrm{V}$, where $\mathrm{V}$ is the active volume, has been around for a while. 
In quartz resonators, this appears either directly or as a side effect of the larger size at lower frequency \cite{Kroupa-1988,Kroupa-2005,vanderZiel-1988,Walls-1992,Driscoll-1993,Sthal-2013}.  
In ultrastable Fabry-Perot cavities, flicker is powered by thermal noise and proportional to the reciprocal of the length \cite{Saulson-1990,Numata-2004} which is approximately equivalent to $1/\mathrm{V}$ after mechanical design rules.

The $1/\mathrm{V}$ law results from a gedankenexperiment in which we combine $m$ equal and independent devices, giving $\mathsf{b}_{-1}|_\text{total}=\mathsf{b}_{-1}|_\text{dev}/m$.  
This has been confirmed experimentally with amplifiers \cite[Chapter\,2]{Rubiola-2008-Leeson}, \cite{Boudot-2012}.  
Flicker is of microscopic origin because the probability density function is Gaussian, which originates from a large statistically-independent population through the central limit theorem.
So, the $m$ devices can be combined in a factor-of-$m$ larger device exhibiting a factor-of-$1/m$ lower flicker.  Similarly, we expect higher flicker if the size of the device is scaled down, until space correlation appears.  The limit for small volume is not known.

In digital electronics, the volume $V$ of the active region is proportional to the node size $\mathcal{S}$.  For reference, $\mathcal{S}$ is of 10 $\mu$m in Intel 4004 (1971), and of 16 nm in the Apple A10 Fusion chip of the iPhone 7.  
While the footprint surface is proportional to $\mathcal{S}^2$, 
the two scaling rules are common in the literature on VLSI systems, known as \emph{constant-voltage} and \emph{Dennard} \cite[P.\,253]{Weste-2011}, \cite{Dennard-1974}, agree in the depth proportional to $\mathcal{S}$.  Thus, $V\propto\mathcal{S}^3$\@.  
The wire delay may contain $\sqrt{\mathcal{S}}$, however, the flicker associated to wires is too small to deserve attention \cite{Verbruggen-1989}.

\begin{figure}
\centering
\includegraphics[width=0.91\columnwidth]{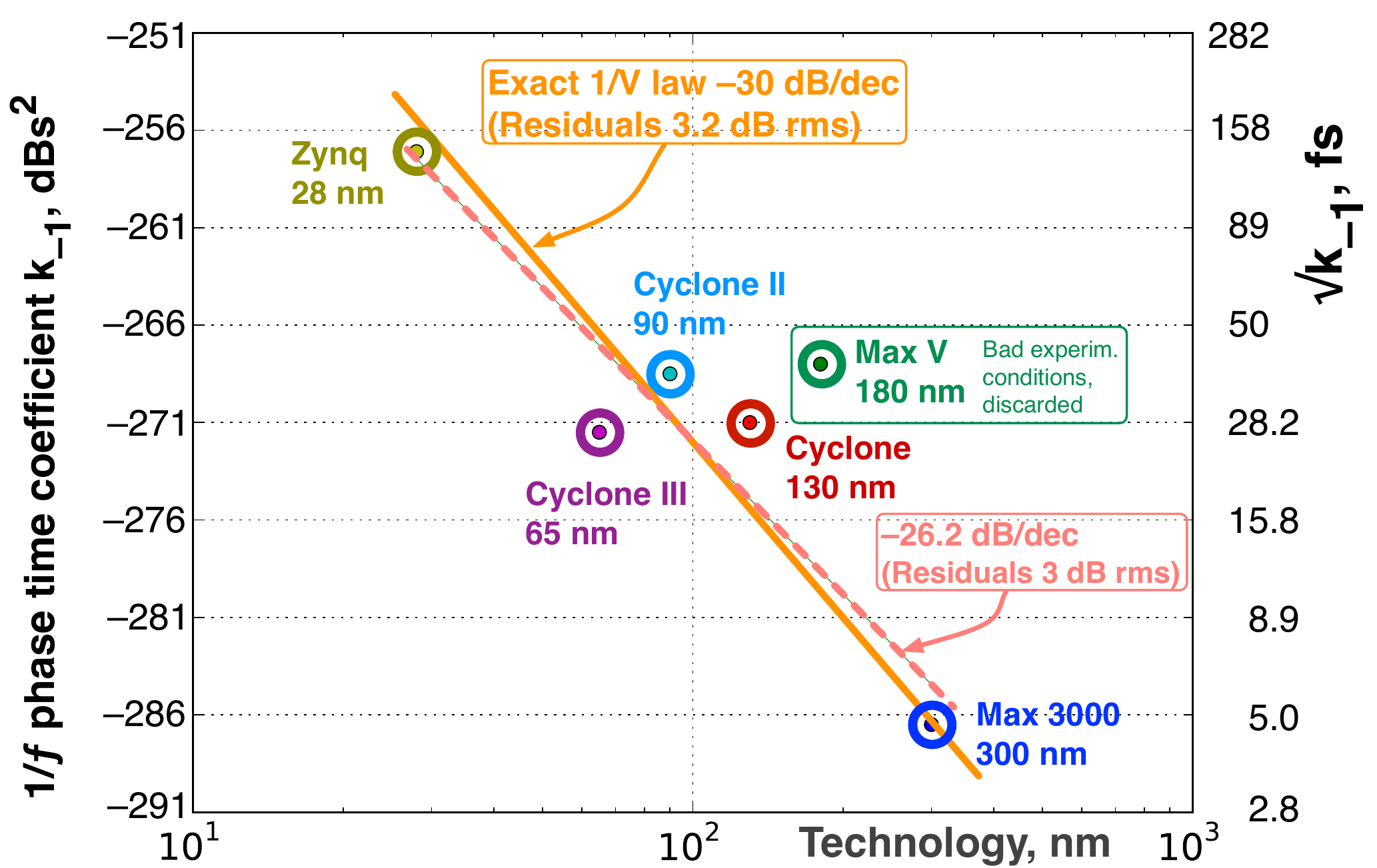}
\caption{Flicker coefficient $\mathsf{b}_{-1}$ of digital devices, related to the cell size $\mathcal{S}$.}
\label{fig:Volume-law}
\end{figure}
We measured a few components using the $\div10$ $\Lambda$ divider configuration.  This gives access to the $1/f$ noise of the clock distribution, which is of the $\mathsf{x}$-type.
We used $100\rightarrow10$ MHz, or $30\rightarrow3$ MHz with the Cyclone and the Cyclone II for practical reasons, sharing a 5125A and a 5120A\@. 
The results are shown in Fig.\,\ref{fig:Volume-law}, which compares the $1/f$ PM noise to $\mathcal{S}$. 

The MAX V is not accounted for in the analysis because the spectrum was taken in unfavorable conditions, yet kept for completeness.
A linear regression gives $\mathsf{k}_{-1}=-26.2 \log_{10}(\mathcal{S}) -219.5$ \unit{dBs^2}, with $\mathcal{S}$ in nm.
Fitting the same data with the exact volume law gives $\mathsf{k}_{-1}=-30 \log_{10}(\mathcal{S})-212.1$ \unit{dBrad^2/Hz}.
The $-26.2$ dB/dec slope is reasonably close to the $1/V$ law ($-30$ dB/dec), with a number of measurement and accuracy insufficient to assess a discrepancy.

\section{Input Chatter}\label{sec:Chatter}
\begin{figure}
\centering
\includegraphics[width=0.71\columnwidth]{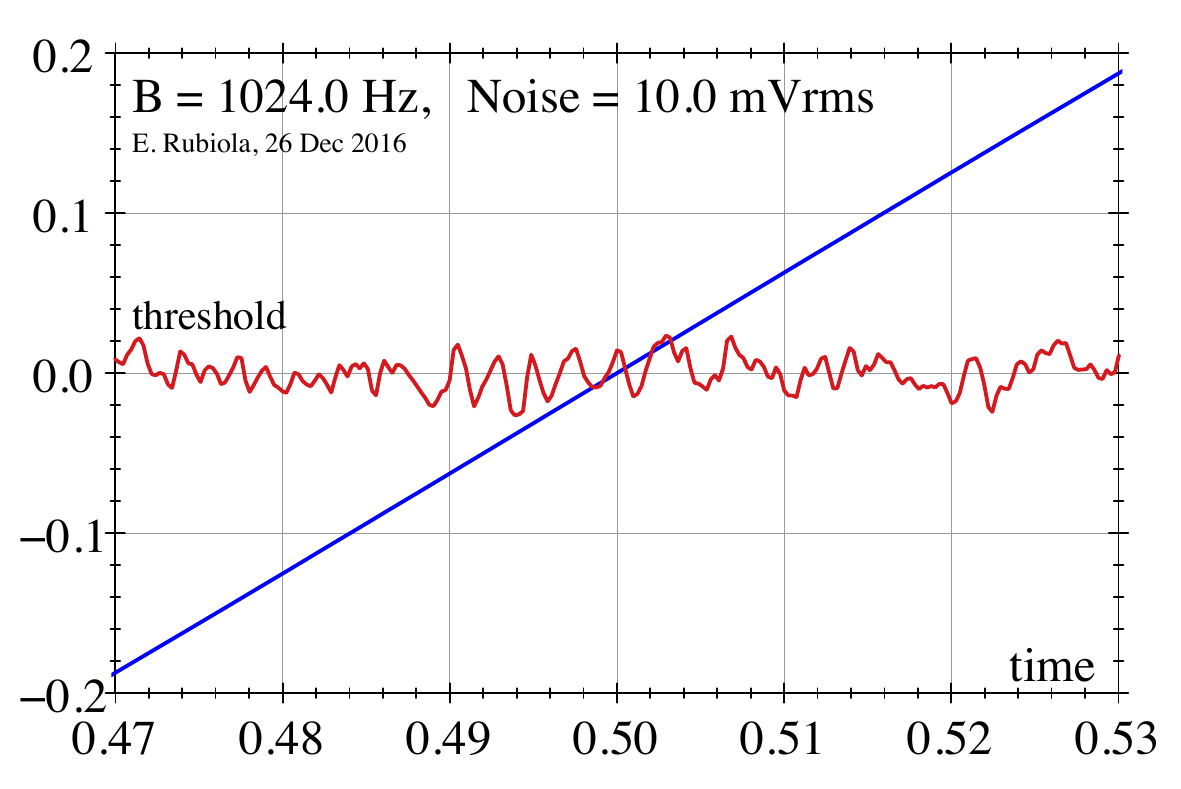}
\caption{Simulation of carrier crossing a fluctuating threshold (normalized 1 Hz carrier, 1 \unit{V_\text{peak}}).  Multiple crossing occurs in the center of the plot.}
\label{fig:Chatter-simulation}
\end{figure}
\begin{figure}
\centering
\includegraphics[width=0.84\columnwidth]{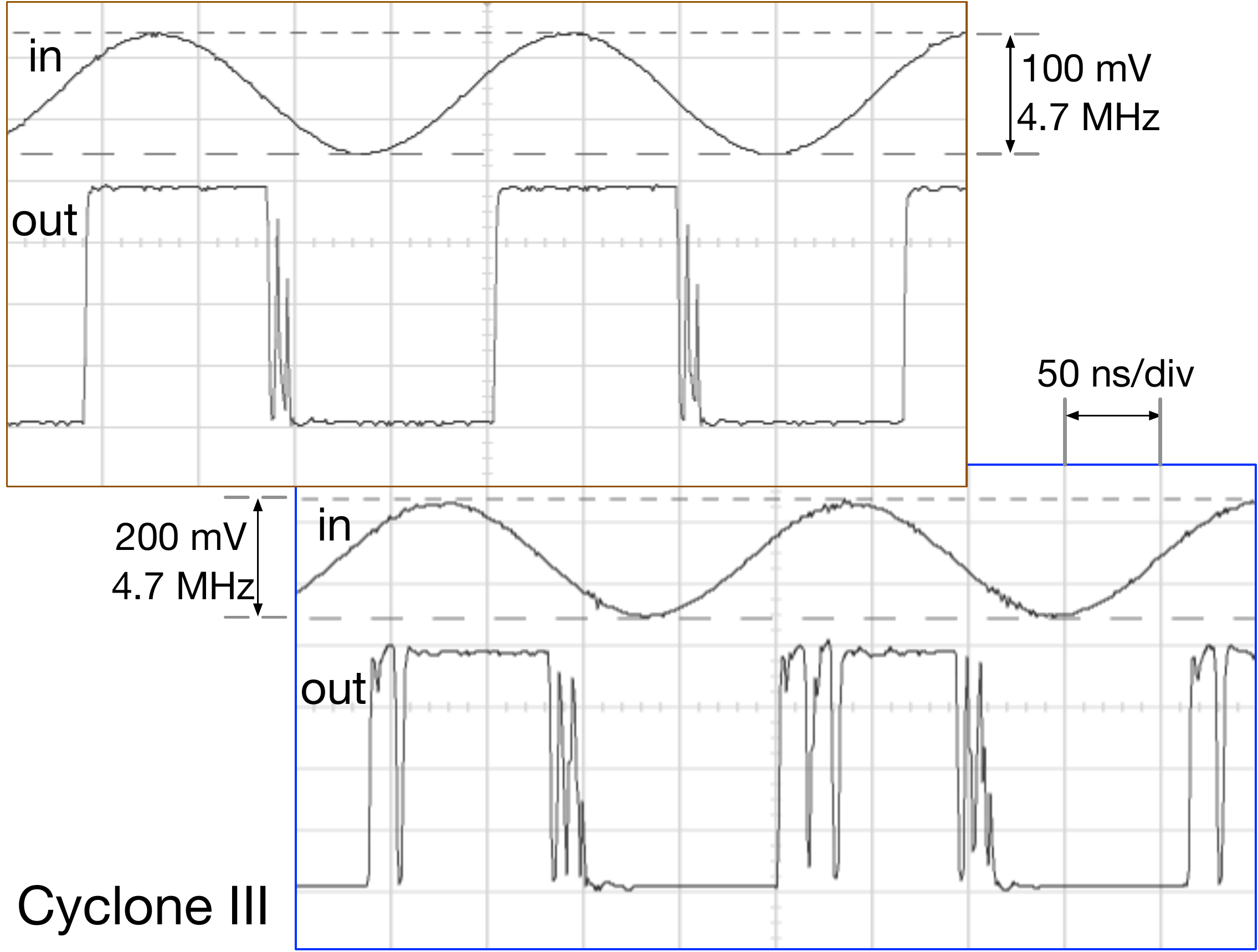}
\caption{Example of chatter (multiple bouncing) when the input SR is insufficient as compared to the SR associated to noise.}
\label{fig:Chatter-example}
\end{figure}
Chatter is a fast random switching of a comparator, which occurs in the presence of wideband noise when the mean square slew rate of noise exceeds that of the signal at the threshold, i.e., $\left<\mathrm{SR}_n^2\right>>\mathrm{SR}_v^2$.  The phenomenon is shown in Fig.\,\ref{fig:Chatter-simulation} and \ref{fig:Chatter-example}.  

Following the Rice's approach
\cite{Rice-1944-BSTJ,Rice-1945-BSTJ},
noise in the small interval $[f, f+\Delta f]$ can be represented as the sinusoidal signal $n_f(t)=V_f\cos(2\pi ft+\theta_f)$, which has random amplitude $V_f$, random phase $\theta_f$, and slew rate
\begin{gather}
\mathrm{SR}_{n,f}=2\pi fV_f \sin(\theta_f)\,.
\label{eqn:Rice-SR}
\end{gather}
The Parseval theorem requires that $\bigl<n^2_f(t)\bigr>=S_{n}(f)\:\Delta f$, thus
\begin{gather}
\bigl<V^2_f\bigr>=2S_n(f)\:\Delta f
\label{eqn:V_f}
\end{gather}
because $\bigl<\cos^2(\dots)\bigr>=1/2$ in $n_f(t)$.  The mean square slew rate
is calculated combining \req{eqn:Rice-SR} and 
\req{eqn:V_f}, integrating on frequency, and averaging on $\theta_f$.  Since $\bigl<\sin^2(\theta_f)\bigr>=1/2$, 
\begin{gather}
\bigl<\mathrm{SR}_n^2\bigr>=4\pi^2\int_0^\infty f^2S_n(f)\:df~.
\label{eqn:Chatter-SR-noise-1}
\end{gather}
In turn, $\bigl<\mathrm{SR}_n^2\bigr>$ is determined by white noise $S_n(f)=\mathrm{h}_0$, $f=[0,B]$.  Other noise types are negligible because they occur al low frequency, compared to $B$, and because of the $f^2$ term in \req{eqn:Chatter-SR-noise-1}.  Thus
\begin{gather}
\left<\mathrm{SR}_n^2\right>=\frac{4\pi^2}{3}\mathrm{h}_0 B^3\,.
\label{eqn:Chatter-SR-noise}
\end{gather}
Since the clock signal \req{eqn:clock} has slew rate $\mathrm{SR}_v=2\pi\nu_0V_0$, the chatter threshold is 
\begin{align}
\nu_0V_0&= \sqrt{\frac{1}{3}\mathrm{h}_0B^3}&&\text{(chatter threshold)}
\label{eqn:Chatter-threshold}\,.
\end{align}

Taking the Cyclone III parameters (Sec.\,\ref{Cyclone-III}, $B=2.5$ GHz and $e_n=11$ \unit{nV/\sqrt{Hz}}, thus $\mathrm{h}_0=1.21{\times}10^{-16}$~\unit{V^2/Hz}), and  $\nu_0=4.7$ MHz, \req{eqn:Chatter-threshold} suggests a threshold $V_0=169$~mV\@.  On Fig.\,\ref{fig:Chatter-example}, we see that chattering occurs at $V_0=100$~mV, and at $V_0=50$~mV the transitions are broken.
Given the difficulty of identifying the parameters,  the agreement between model and observation is satisfactory.

After \req{eqn:Chatter-threshold}, chattering is more likely at low carrier frequency.  However, Fig.\,\ref{fig:Chatter-example} shows that this can occur at 5 MHz, a standard frequency of great interest for high stability signals.

\section{Internal PLL}\label{sec:PLL}
\begin{figure}
\centering\includegraphics[scale=0.64]{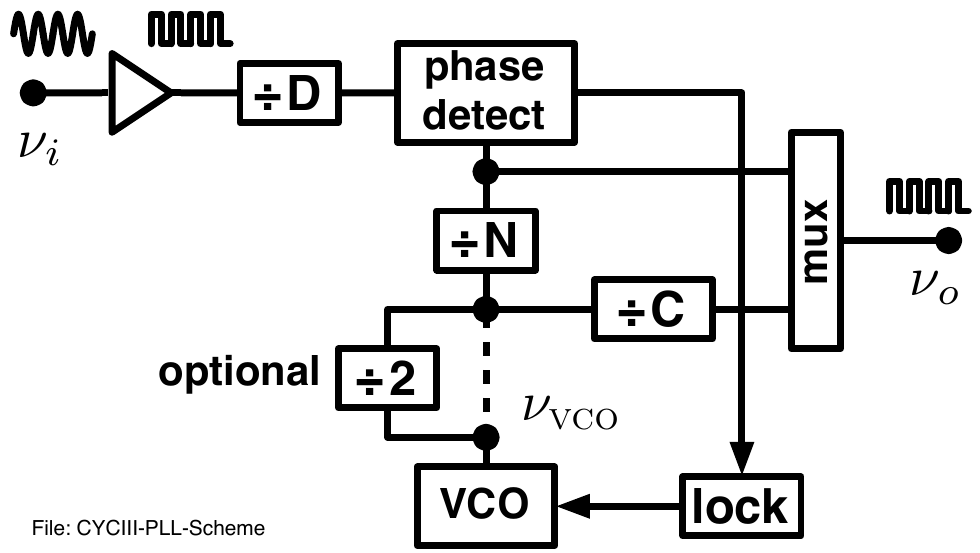}
\caption{Cyclone III internal PLL frequency multiplier.}
\label{fig:CYCIII-PLL-Scheme}
\end{figure}

The internal PLL is intended to provide high frequency internal clock stabilized to an external reference, often 5-10-100 MHz.
We show simple experiments which give insight in the Cyclone~III.

The PLLs is shown in Fig.\,\ref{fig:CYCIII-PLL-Scheme}.  The VCO operates in the 0.6--1.3 GHz range, extended to 300--650 MHz by the optional $\div2$ divider, always present in our tests.
A classical phase-frequency detector (PFD) is present, with charge pump output driving the analog feedback to the VCO\@.
The PLL output frequency is $\nu_o=\frac{N}{CD}\nu_i$.  This leaves three degrees of freedom ($N$, $C$ and $D$), two of which are available to the designer.  The programming tool (Quartus) uses one to ensure that internal design rules are satisfied.

The VCO relies on a LC resonator on chip.  General literature suggests a quality factor $Q$ of 5--10, limited by the technology \cite{Hajimiri-1999}.  Therefore, we expect a Leeson frequency $f_L=\nu_\textsc{vco}/2Q$ of the order of 50 MHz.

\begin{figure}[t]
\centering\includegraphics[width=0.84\columnwidth]{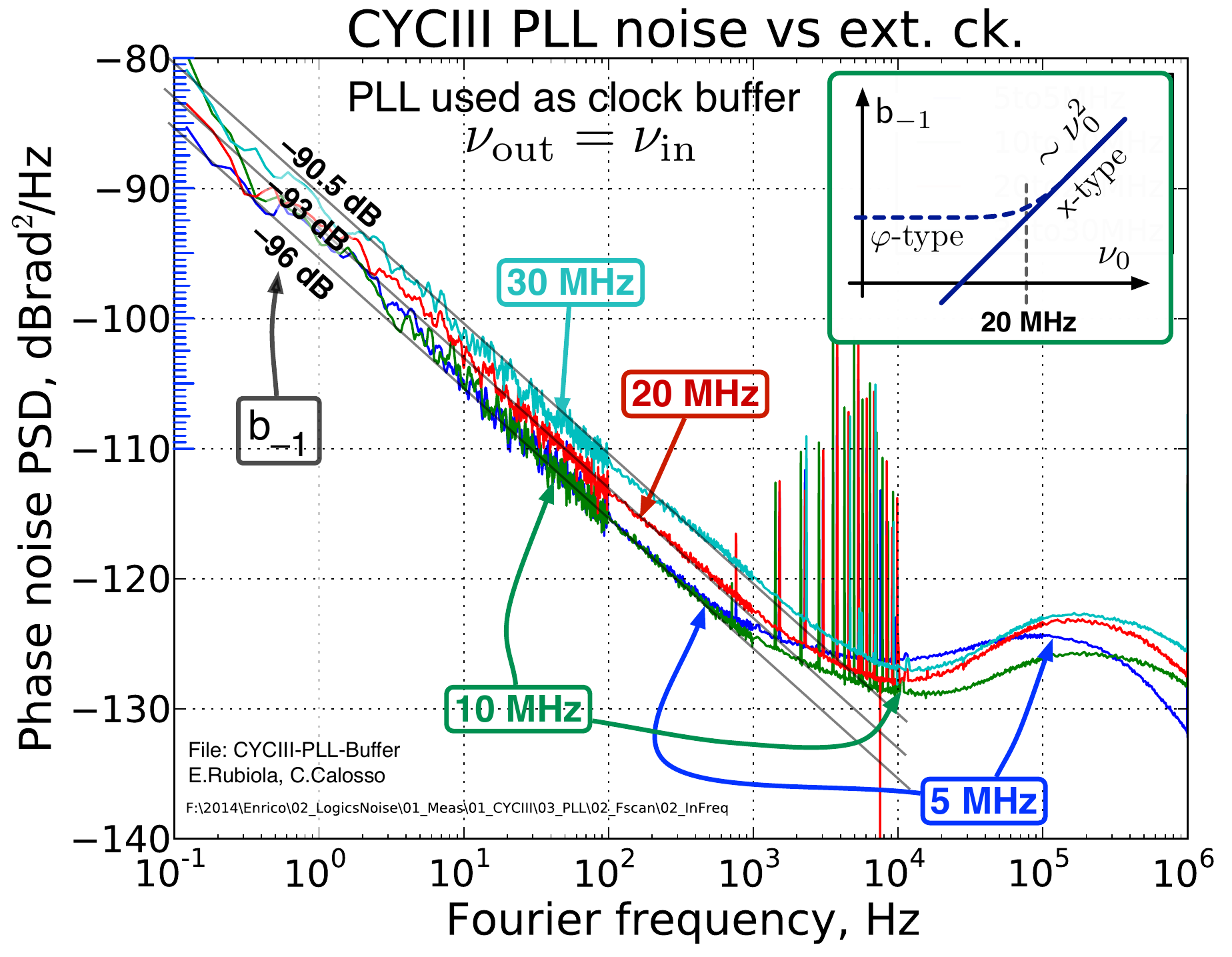}
\caption{The internal PLL is used as a buffer, that is, $\nu_o=\nu_i$.}
\label{fig:CYCIII-PLL-Buffer}
\end{figure}
In a first experiment (Fig.\,\ref{fig:CYCIII-PLL-Buffer}), we use the PLL as a `cleanup' ($\nu_o=\nu_i$), yet with a high purity input.   This gives the noise of the PLL, at different values of $\nu_i$.  
For lowest noise, we use the phase comparator at the highest possible frequency ($\nu_i$) by setting $D=1$.
The VCO frequency ends up to be 400, 600 or 640 MHz, depending on $\nu_o$.
On Fig.\,\ref{fig:CYCIII-PLL-Buffer}, the white noise floor is not seen.  This is sound because noise can be white only beyond $f_L$, which is beyond the 1 MHz span.
Flicker is of the $\varphi$-type at 5 and 10 MHz, with $\mathsf{b}_{-1}=2.5{\times}10^{-10}$ \unit{rad^2/Hz} ($-96$ dB).  Since this type of noise is \emph{not} scaled down by the $\div N$ divider in the loop, we ascribe it to the \emph{phase detector}. 
This is because (i) with the tight lock implemented we do not expect to see the VCO; and (ii) the input comparator and the output stage of the $\div N$ divider have some 10 dB lower noise in similar conditions ($-115$ \unit{dBrad^2/Hz}, Section~\ref{Cyclone-III}).

\begin{figure}[t]
\centering\includegraphics[width=0.84\columnwidth]{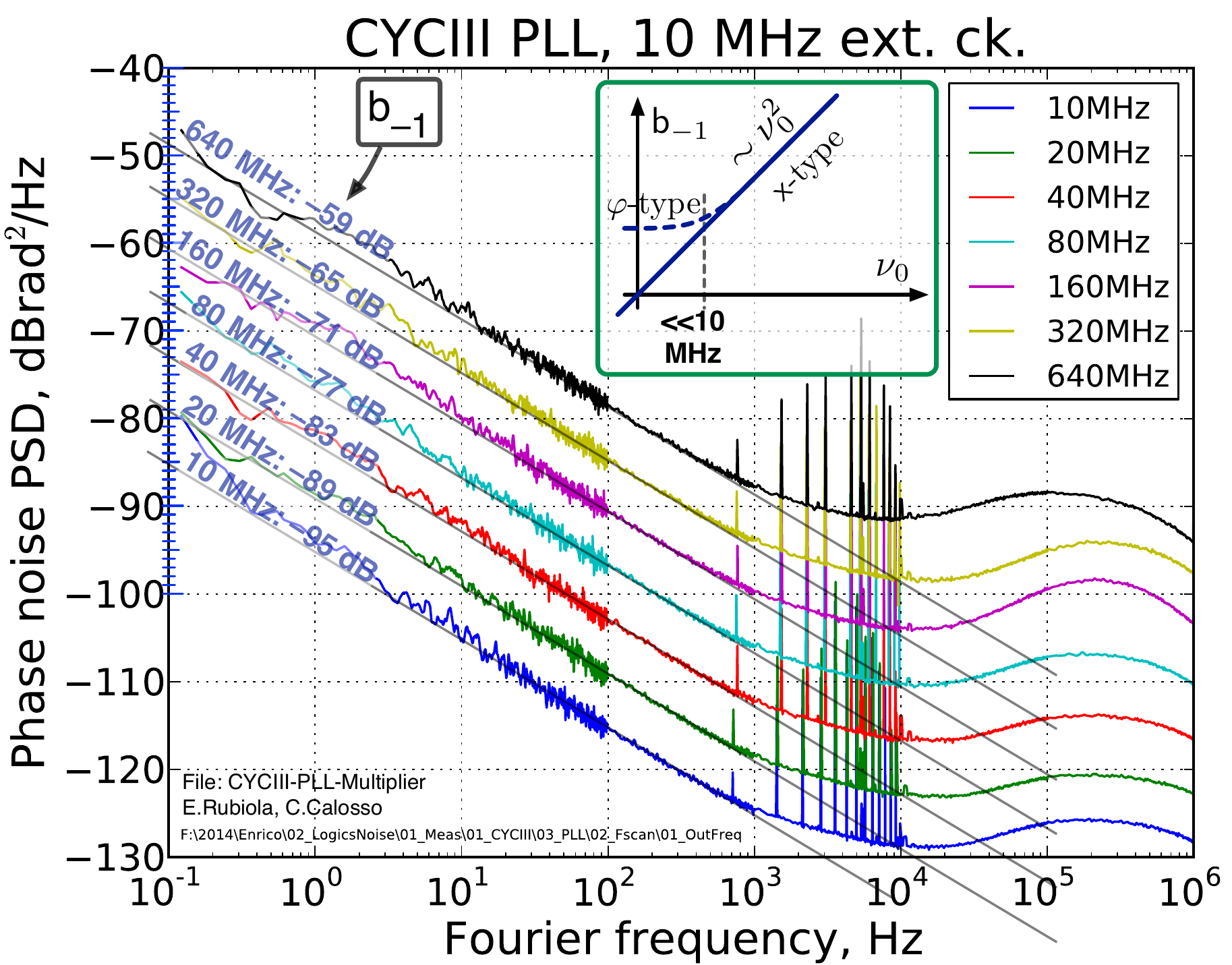}
\caption{The internal PLL is used as a frequency multiplier in powers-of-two of multiples of the 10 MHz frequency reference.}
\label{fig:CYCIII-PLL-Multiplier}
\end{figure}
In the second experiment, we use the PLL as a frequency multiplier in powers of two ($\nu_o=2^m\nu_i$) from 10 MHz to 640 MHz, with $\nu_i=10$ MHz.
Again, we use $D=1$ for lowest noise.  
The VCO delivers 320, 400 or 640 MHz, depending on $\nu_o$.
The phase noise spectrum (Fig.\,\ref{fig:CYCIII-PLL-Multiplier})
indicates that flicker is of the $\mathsf{x}$-type, scaling up as $\nu_o^2$.  This indicates that the phase detector is the dominant source of noise, with negligible contribution of the dividers.  So, the time fluctuation $\mathsf{x}(t)$ is transferred from the phase detector to the VCO, and then from the VCO to the output.  The phase $\varphi(t)$ scales accordingly, that is, $\times N/C$.

\section{Thermal Effects}\label{sec:Thermal-effects}

\subsection{Thermal Transients}\label{sec:Thermal-transients}
Common sense suggests that delay is affected by the junction temperature $T_J$, while other parameters like $T_C$ and $T_A$ (case and ambient temperature) are comparatively smaller importance.

Our method consists in using the electrical power $P$ to heat the chip, and calculate $T_J$ from the thermal resistance $\Theta_{JA}$ and the transients.
In turn, $P$ is chiefly set by the charge/discharge cycle of the gate capacitance, whose energy is $E=CV^2$.  Thus, $N$ gates switching at $\nu_0$ dissipate $P=NCV^2\nu_0$. Of course, $P$ can be changed instantaneously.
The delay is measured with a Symmetricom 5125A test set used as a phase meter and also as a time-interval counter.

\begin{figure}[t]
\centering\includegraphics[width=0.91\columnwidth]{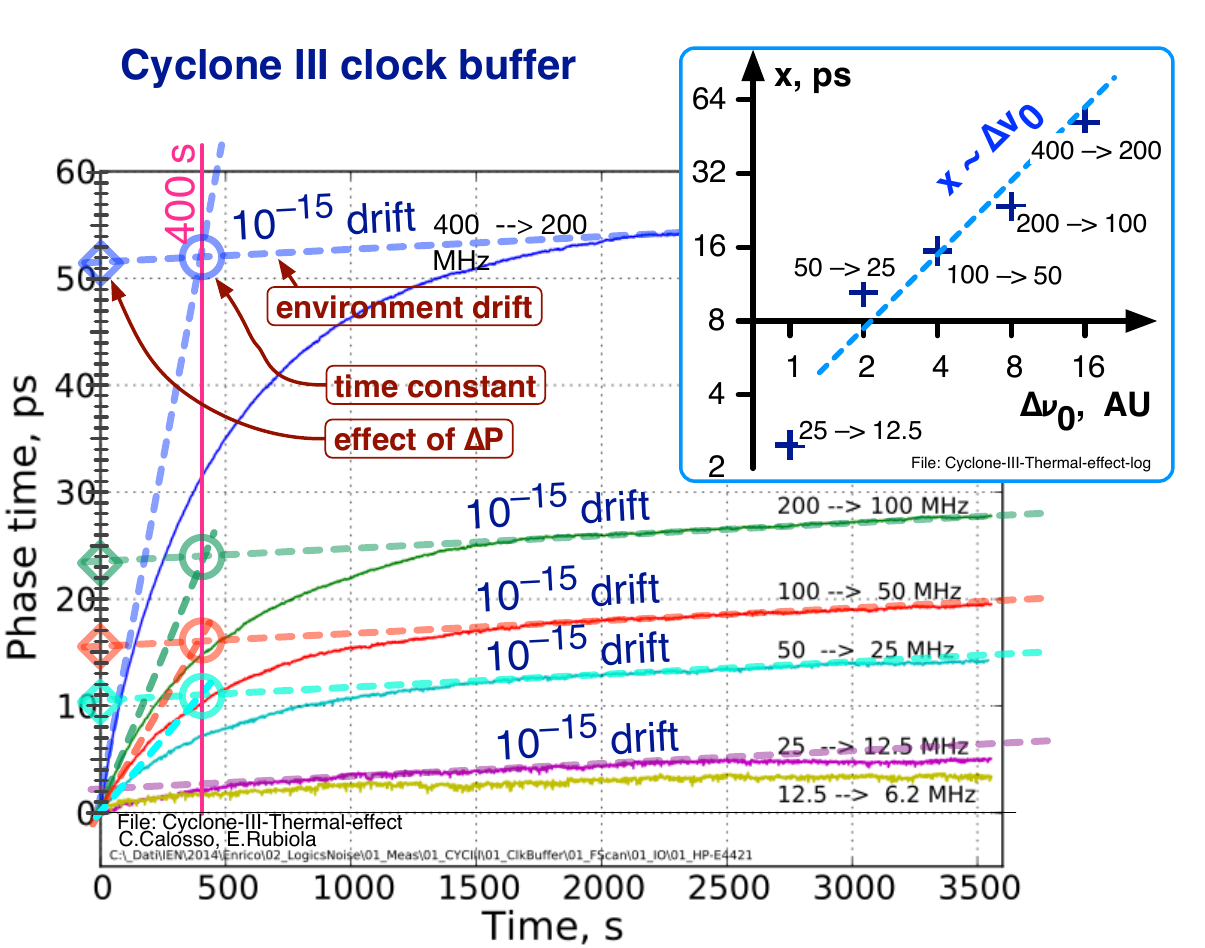}
\caption{Thermal effects measured on a Cyclone III FPGA\@.  Each curve represents the thermal transient when the clock frequency is divided by two.}
\label{fig:Cyclone-III-Thermal-effect}
\end{figure}

We measured a Cyclone~III used as a clock buffer (actually, 10 buffers connected in parallel through 330~\ohm\ resistors).  The temperature had to be low-pass filtered by covering the card with a small piece of tissue.
The results are shown in Fig.\,\ref{fig:Cyclone-III-Thermal-effect}.

In the main body, all the curves show an exponential behavior plus a linear drift
\begin{gather}
\mathsf{x}(t)=k'\Delta T\big(1-e^{-t/\tilde{\tau}}\big)+k''t~,
\end{gather}
where $\Delta T=T_J-T_A$ results from setting $\nu_0$ in powers of two, and $\tilde{\tau}$ is the time constant.  For reference, we observed $P=1$ W at 400 MHz, which means $\Delta T\approx10$ K with $\Theta_{JA}\approx10$ K/W (including the thermal pad on the pcb), and neglecting the dissipation at $\nu_0=0$.

The linear drift (1 fs/s, or $10^{-15}$ fractional frequency) does not scale with power.  This behavior is typical of the environment temperature, slowly drifting during the measurement (a fraction of a Kelvin over 1 hour).
Extrapolating the drift to $t=0$, we get the asymptotic effect of the $\Delta P$ transient alone.  

The time constant $\tilde{\tau}$ is found as the intercept of the tangent at $t=0$ and the linear drift (dashed lines).  This graphical process removes the drift.  The value $\tilde{\tau}=400$ s is  the same for all the transients.

The inset of Fig.\,\ref{fig:Cyclone-III-Thermal-effect} shows the delay versus the carrier frequency (dissipated power).  As expected, the delay is proportional to $T_J$, set through $\nu_0$.  
Accounting for $P$ and $\Theta_{JA}$, the thermal coefficient of the delay is $10$ ps/K\@.

\subsection{Allan Deviation}
\begin{figure}
\centering
\includegraphics[width=\columnwidth]{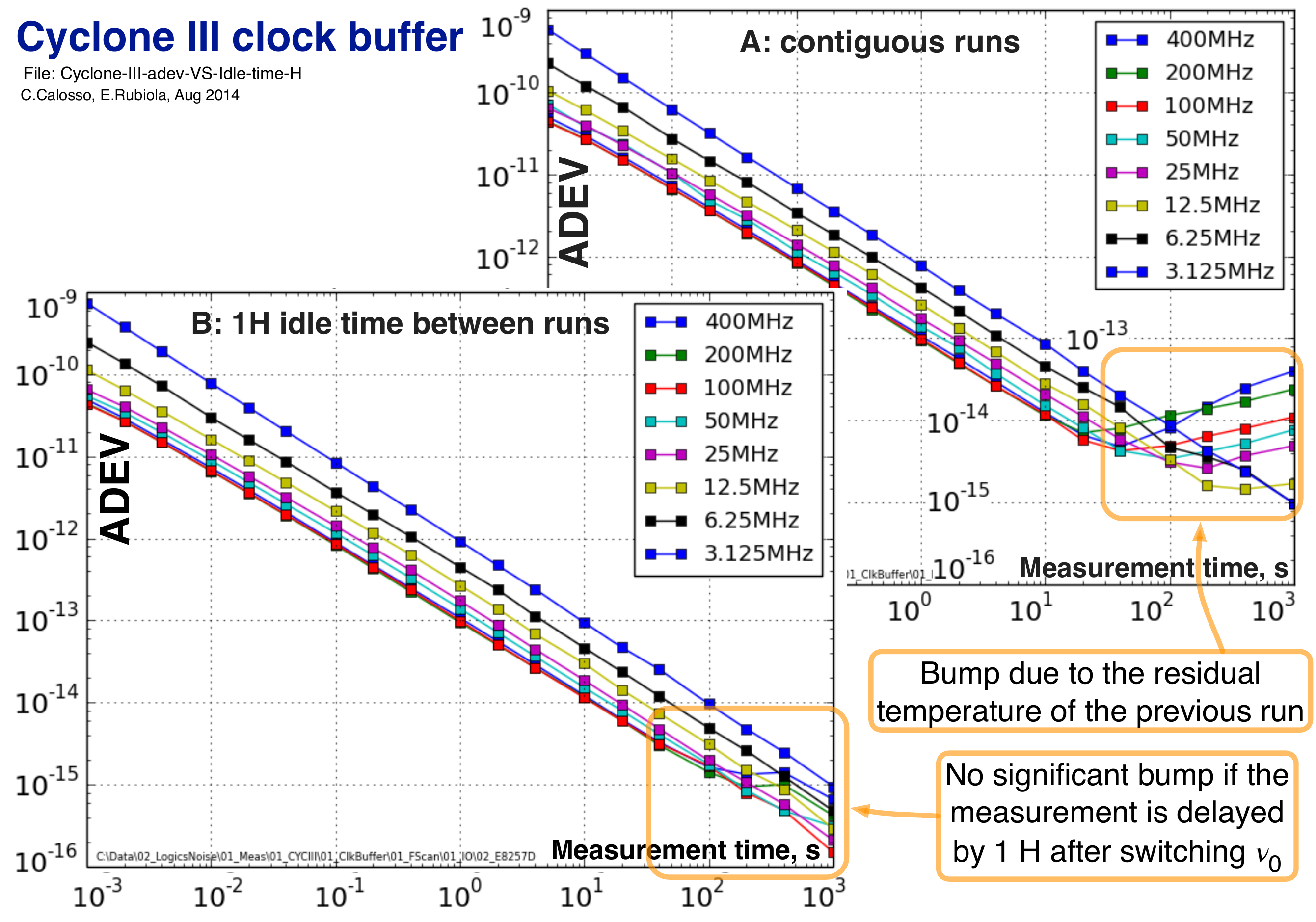}
\caption{Allan deviation $\sigma_\mathsf{y}(\tau)$ derived from the FPGA delay.}
\label{fig:Cyclone-III-adev-VS-dead-time}
\end{figure}

Generally, $\sigma_\mathsf{y}(\tau)$ should follow the $1/\tau$ law (white and $1/f$ phase noise).  Other types of instability, as frequency noise would reveal a phase noise steeper than $1/f$, and the delay of the device would diverge in the long run.  However, bumps may be present.
Notice that $1/f$ phase noise in practice never yields large integrated delay. 

Figure~\ref{fig:Cyclone-III-adev-VS-dead-time} shows
the Cyclone III Allan deviation $\sigma_\mathsf{y}(\tau)$, measured with a Symmetricom 5125A test set.

We first discuss the $1/\tau$ region of Fig.\,\ref{fig:Cyclone-III-adev-VS-dead-time}\,A\@.  
At low $\nu_0$, $\sigma_\mathsf{y}(\tau)$ decreases proportionally to $1/\nu_0$.  
For $\tau=1$ s, we read $\sigma_\mathsf{y}=10^{-12}$ at 3.125 MHz, $5{\times}10^{-13}$ at 6.25 MHz, etc. 
At higher $\nu_0$ the curves get closer to one another, and overlap at $\nu_0\ge100$ MHz.

Taking the classical conversion formulae for Allan variance and spectra (for example, \cite[P.\,77--80]{Kroupa-1983}, or \cite{IEEE-STD-1139-2008}), 
the $1/\nu_0$ behavior is equivalent to $\mathsf{h}_{1}\propto1/\nu_0^2$ (frequency fluctuation spectrum $S_\mathsf{y}(f)=\mathsf{h}_{1}f$), thus to $\mathsf{b}_{-1}=C$ vs.\ $\nu_0$.  
This is the signature of the pure $\varphi$-type noise, as expected at low $\nu_0$ and at low $f$, thus at long $\tau$.
We recall that the fluctuation of the input threshold is dominant at low $\nu_0$, and that the low $f$ region is dominated by the $1/f$ phase noise, virtually unaffected by aliasing.

By contrast, the $\sigma_\mathsf{y}(\tau)=C$ vs.\ $\nu_0$ behavior is equivalent to $\mathsf{h}_{1}=C$ vs.\ $\nu_0^2$, thus $\mathsf{b}_{-1}\sim\nu_0^2$.  This is the typical of the pure $\mathsf{x}$-type noise, as expected at high $\nu_0$ and at low $f$, thus at long $\tau$.
The fluctuation of the input threshold is no longer relevant, and the low $f$ region is still dominated by the $1/f$ phase noise, virtually unaffected by aliasing.

In summary, the $1/\tau$ region of the $\sigma_\mathsf{y}(\tau)$ plot is consistent with the predictions of Section~\ref{sec:Noise-model}.

On the right hand of Fig.\,\ref{fig:Cyclone-III-adev-VS-dead-time}\,A, $\sigma_\mathsf{y}(\tau)$ seems to leave the $1/\tau$ law.  This can only be a local phenomenon, i.e. a bump.
Carrying on the experiment, in Fig.\,\ref{fig:Cyclone-III-adev-VS-dead-time}\,A the measurement of $\sigma_\mathsf{y}(\tau)$ restarts immediately after switching $\nu_0$, while in Fig.\,\ref{fig:Cyclone-III-adev-VS-dead-time}\,B the measurement of $\sigma_\mathsf{y}(\tau)$ is delayed by 1 hour after switching $\nu_0$.  
The relevant difference is that in A each curve suffers from the cooling-down transient of the previous measurement, while in B each measurement starts in steady state.
Bumps show up in A at $\tau\ge30$ s, and they get stronger at higher $\nu_0$, where the thermal dissipation is stronger, and almost disappear in B\@.  This is a qualitative confirmation of the presence of two separate time constants (end of Sec.\,\ref{sec:Thermal-transients}).

\subsection{Side Effects of the Thermal Dissipation}
We have shown that the electrical activity inside the FPGA heats the chip, and in turn affects the delay.  Variations exceeding 50 ps have been observed in the presence of a light burden.    
The analysis gives a warning, thermal crosstalk is around the corner when the same FPGA is in charge of more than one task, made worse by the heat latency. 
Attempts to fit low noise and high-stability functions (frequency dividers, etc.) in a chip processing at high rate may be difficult or give unpredictable results.

\section*{Acknowlegments}\label{sec:Acknow}
This work is a part of the ``Programme d'Investissement d'Avenir'' projects in progress in Besancon, i.e., Oscillator IMP, First-TF, and Refimeve+.  Funds come from the ANR, the Region Franche Comt\'{e}, INRIM, and EMRP Project IND 55 Mclocks.

We thank the Go Digital Working Group for general help and fruitful discussion, and among them chiefly Jean-Michel Friedt, Pierre-Yves ``PYB'' Bourgeois, and Gwenhael ``Gwen'' Goavec-M\'{e}rou.

\def\bibfile#1{/Users/rubiola/CloudStation/Drive/Author/!-Bibliography/#1}
\bibliographystyle{IEEEtran}
\bibliography{\bibfile{ref-short},%
              \bibfile{references},%
              \bibfile{rubiola},%
              Local}

\end{document}